\documentclass[
prc,%
10pt,%
final,%
notitlepage,%
oneside,%
twocolumn,%
nobibnotes,%
nofootinbib,
superscriptaddress,%
floatfix,%
showkeys,%
showpacs]%
{revtex4}
\usepackage{color}
\usepackage{amsfonts}
\usepackage{amsbsy}
\usepackage{mathrsfs}
\usepackage{graphicx}
\def\lsim{\mathrel{\rlap{
\lower4pt\hbox{\hskip-3pt$\sim$}}
    \raise1pt\hbox{$<$}}}     
\def\gsim{\mathrel{\rlap{
\lower4pt\hbox{\hskip-3pt$\sim$}}
    \raise1pt\hbox{$>$}}}     
\def\scr#1{\mbox{\scriptsize #1}}
\begin{document}
\title{Phase Evolution and Freeze-out within Alternative Scenarios of Relativistic Heavy-Ion Collisions}%
\author{Yu.~B.~Ivanov}\thanks{e-mail: Y.Ivanov@gsi.de}
\affiliation{Kurchatov Institute, 
Moscow RU-123182, Russia}
\begin{abstract}
Global evolution of the matter in relativistic collisions of heavy nuclei 
and the resulting global freeze-out parameters  
are analyzed in a wide range of incident energies  
 2.7 GeV  $\le \sqrt{s_{NN}}\le$ 39 GeV. The analysis is performed  
within  the three-fluid model
employing three different equations of state (EoS): a purely hadronic EoS,   
an EoS with the first-order phase transition  
and that with a smooth crossover transition.
Global freeze-out parameters deduced from experimental data 
within the statistical model are well reproduced within the crossover scenario. 
The 1st-order-transition scenario is slightly less successful. 
The worst reproduction is found within the purely hadronic scenario. 
These findings make a link between the EoS and results of the statistical model,  
and indicate that deconfinement onset occurs at $\sqrt{s_{NN}}\gsim$ 5 GeV.
\end{abstract}
\maketitle

\section{Introduction}

Extensive simulations of relativistic heavy-ion collisions
were performed within a model of the three-fluid 
dynamics (3FD) \cite{3FD} employing three different equations of state (EoS): a purely hadronic EoS   
\cite{gasEOS} (hadr. EoS), which was used in the major part of the 3FD simulations so far 
\cite{3FD,3FD-hadr}, and two versions of EoS involving the deconfinement 
transition \cite{Toneev06}. These two versions are an EoS with the first-order phase transition
and that with a smooth crossover transition. 
These simulations cover the energy range from 2.7 GeV 
to 39 GeV in terms of center-of-mass energy, $\sqrt{s_{NN}}$.
Details of the calculations are described in Ref.  
\cite{Ivanov:2013wha} dedicated to analysis of the baryon stopping. 
With these EoS's, onset of the deconfinement transition occurs at top AGS energies, 
i.e. $\sqrt{s_{NN}}\gsim$ 5 GeV, as shown in Refs. \cite{Ivanov:2013wha,Ivanov:2012bh}. 
The results \cite{Ivanov:2013wha,Ivanov:2012bh,Ivanov:2013yqa,Ivanov:2013mxa} obtained 
so far indicate preference of  deconfinement-transition scenarios in reproducing the 
available experimental data.

In particular, it was found \cite{Ivanov:2013yqa} that the hadronic scenario fails to reproduce 
experimental yields of antibaryons (strange and nonstrange), 
starting already from lower SPS energies, i.e. $\sqrt{s_{NN}}\ge$ 6.4 GeV, 
and  yields of all other species
at energies above the top SPS one, i.e. $\sqrt{s_{NN}}>$ 17.3 GeV, while 
the  deconfinement-transition scenarios reasonably agree (to a various extent) with all the data. 
It is naturally to search for a reason of this fact  in differences of the final freeze-out 
states produced by different scenarios. Indeed, the statistical model (SM) needs only two
parameters, temperature ($T$) and baryon chemical potential ($\mu_B$), to describe 
ratios of  (total and midrapidity) yields of all the produced species
\cite{BraunMunzinger:1994xr,BraunMunzinger:1995bp,Cleymans:1999st,Becattini:2000jw,%
Becattini:2003wp,Becattini:2005xt,Andronic:2005yp,Andronic:2006ky,Andronic:2008gu}. 
If the 3FD evolution 
drives the system to a final freeze-out state characterized by proper $T$ and $\mu_B$
(somehow averaged over the system), 
then the experimental hadron yields are reproduced. 
Of course, the 3FD freeze-out state is characterized by 3D fields of $T$ and $\mu_B$.  
The $(T,\mu_B)$  point in question is formed by values around which these 
fields are centered.

In fact, the same procedure of the 
freeze-out with the same freeze-out energy density \cite{3FD,71,74} was used 
in all considered scenarios of nuclear collisions. 
Nevertheless, the final states in different scenarios turn out to be different because 
the phase evolution of the system is determined by the specific EoS. 
Of course, these final states are also characterized by fields of collective flows 
rather than only the temperature and baryon chemical potential, and hence 
the 3FD model pretends to describe not only hadron yields. However, for the 
particular case of the hadron yields the position of the final freeze-out state
in the $(T,\mu_B)$ phase space is of prime importance.

Therefore, in this paper I analyze the 3FD final freeze-out state in terms of its position in 
the $(T,\mu_B)$ phase space. This analysis extends to 
relativistic heavy-ion collisions in the energy range from 2.7 GeV 
to 39 GeV in terms of $\sqrt{s_{NN}}$. This domain covers 
the energy range of the beam-energy scan program at the 
Relativistic Heavy-Ion Collider (RHIC) at Brookhaven National Laboratory (BNL),  
low-energy-scan program  at Super Proton Synchrotron (SPS)
at CERN and the Alternating Gradient
Synchrotron (AGS) at BNL, as well as newly constructed 
Facility for Antiproton and Ion Research (FAIR) in Darmstadt and the
Nuclotron-based Ion Collider Facility (NICA) in Dubna. 
\section{Freeze-out in 3FD Model}
\label{Freeze-out}

The 3-fluid approximation is a minimal way to 
simulate the finite stopping power at high incident energies.
Within the 3-fluid approximation 
a generally nonequilibrium distribution of baryon-rich
matter is simulated by counter-streaming baryon-rich fluids 
initially associated with constituent nucleons of the projectile
(p) and target (t) nuclei. In addition, newly produced particles,
populating the mid-rapidity region, are associated with a fireball
(f) fluid.
Each of these fluids is governed by conventional hydrodynamic equations
which contain interaction terms in their right-hand sides. 
These interaction terms describe mutual friction of the fluids and 
production of the fireball fluid.
The friction between fluids was fitted to reproduce
the stopping power observed in proton rapidity distributions for each EoS, 
as it is described in  Ref. \cite{Ivanov:2013wha} in detail.

A conventional way of applying the fluid dynamics to heavy-ion collisions at
RHIC and LHC energies is to prepare the initial state for the hydrodynamics by 
means of various kinetic codes, see, e.g., Refs. \cite{Bleicher08,Bleicher09,Hama:2004rr,Nonaka:2012qw}.
Contrary to these approaches, the 3FD model treats the collision process from the 
very beginning, i.e. the stage of cold nuclei, up to  freeze-out within  
the fluid dynamics. 
Therefore, any tuning of initial conditions is impossible within the 3FD model.   

The freeze-out  is performed accordingly to the procedure described in Ref. \cite{3FD} 
and in more detail in Refs. \cite{71,74}. This is a modified Milekhin version of the 
freeze-out that possesses exact conservation of the energy, momentum and baryon number. 
Contrary to the conventional Cooper--Frye  approach \cite{Cooper}, the modified Milekhin 
method has no problem associated with negative contributions to particle spectra. 
This method of freeze-out can be called dynamical, since the
freeze-out process here is integrated into fluid dynamics. 
This kind of freeze-out is similar to the model of ``continuous
emission'' proposed in Ref. \cite{Sinyukov02}. There the particle emission
occurs from a surface layer of the mean-free-path width. In the 3FD case the
physical pattern is similar, only the mean free path is shrunk to zero.

The freeze-out criterion is $\varepsilon < \varepsilon_{\scr{frz}}$, 
where $\varepsilon$
is the total energy density of all three fluids in the proper reference
frame, where the composed matter is at rest.
The freeze-out energy density $\varepsilon_{\scr frz}=$ 0.4 GeV/fm$^3$ 
was chosen mostly on the condition of the best reproduction 
of secondary particles yields (more precisely, mid-rapidity pion densities) 
for all considered scenarios.  However, the freeze-out
front is not defined just ``geometrically'' on the condition of the
freeze-out criterion met but rather is a subject the fluid evolution.
It competes with the fluid flow and not always reaches the place where
the freeze-out criterion is first met.
Therefore, $\varepsilon_{\scr{frz}}$ can be called a "trigger" value of
the freeze-out energy density, whereas the actual thermodynamical parameters 
of the frozen out matter are jointly determined by this "trigger" value and the fluid dynamics
and thus depend on the EoS.

\begin{figure}[bht]
\includegraphics[width=6.2cm]{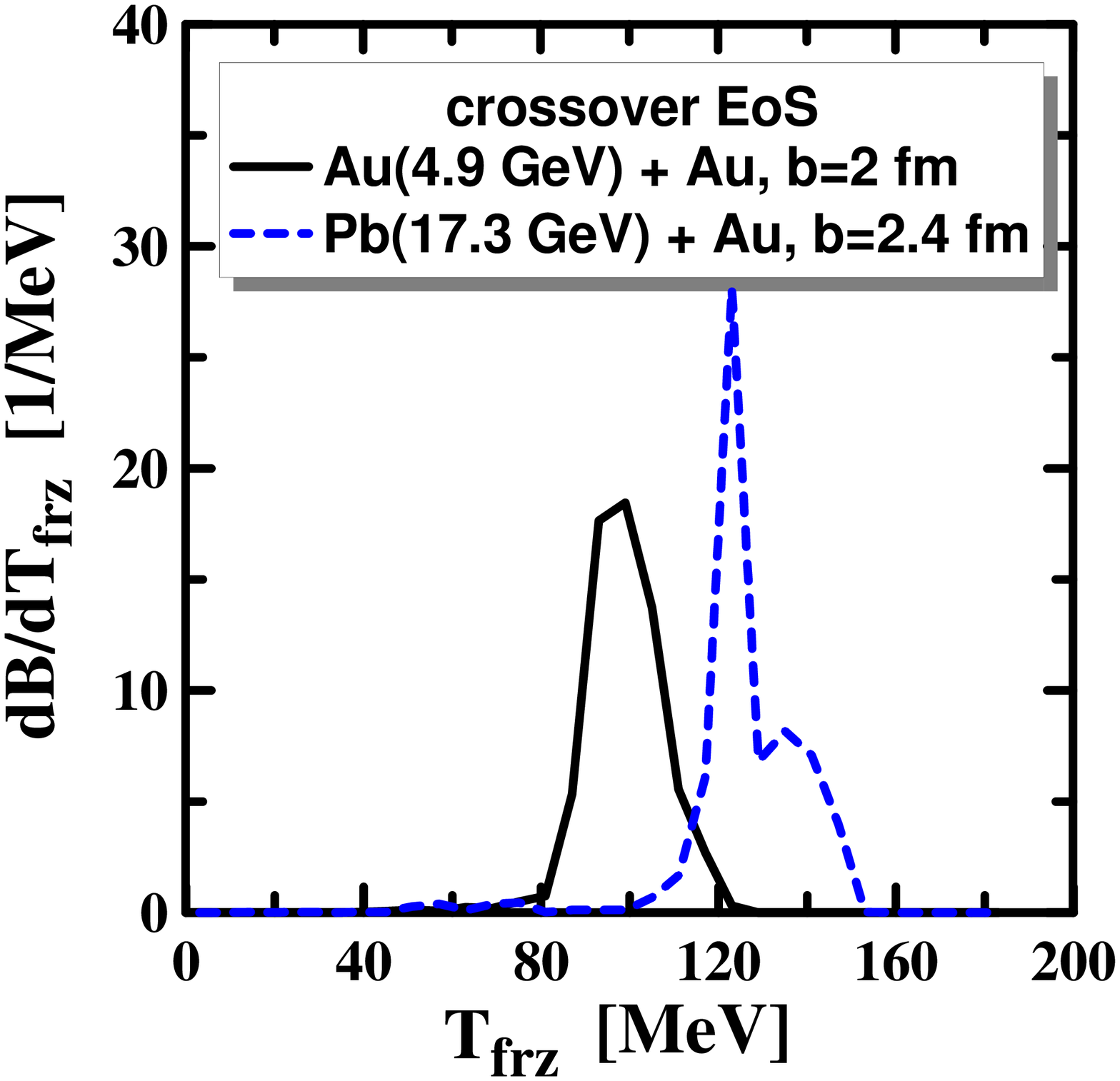}
\\
\includegraphics[width=5.9cm]{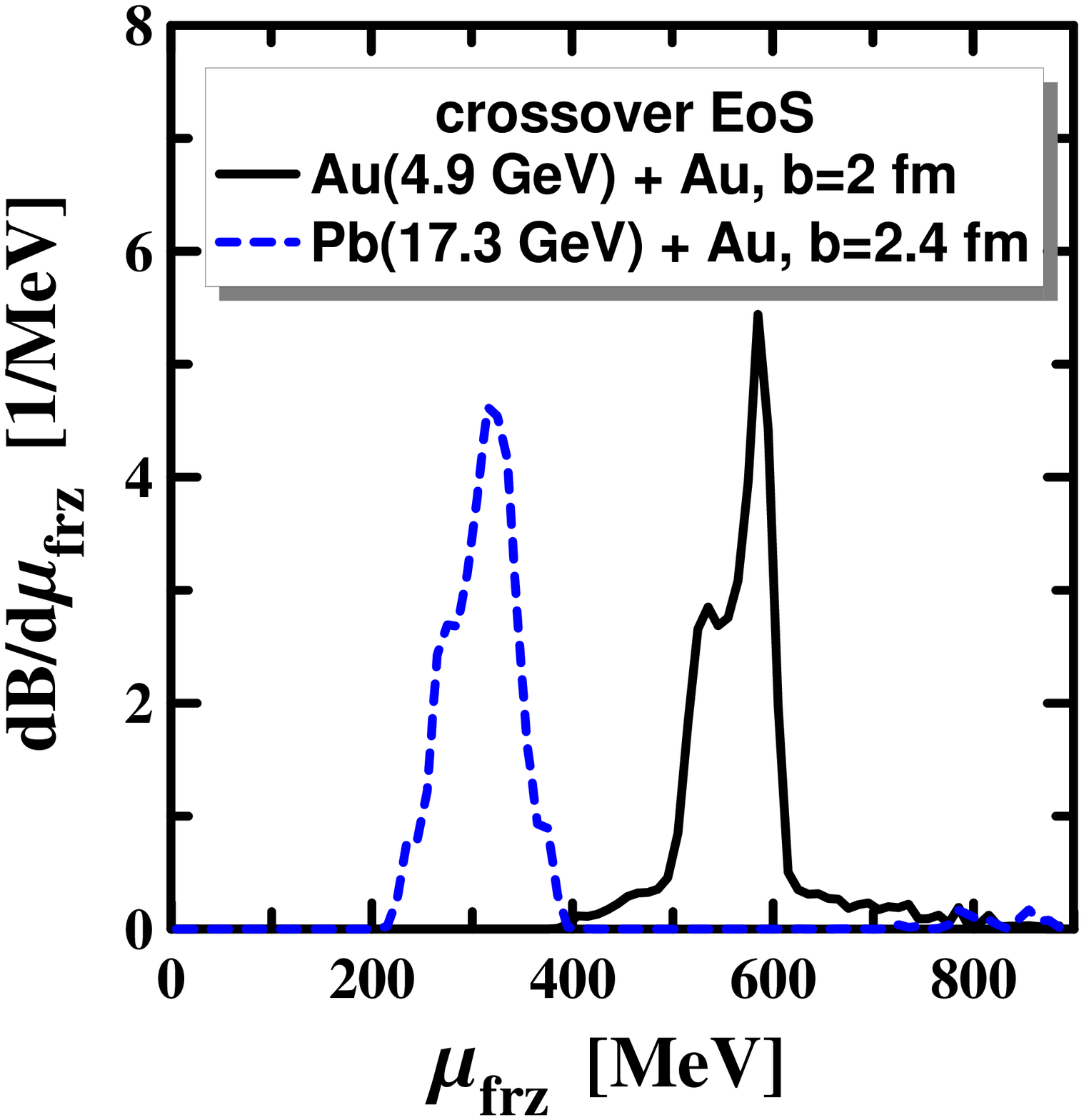}
 \caption{
Distributions of the frozen-out baryon charge over 
temperature (upper panel)
and baryon chemical potential (lower panel)
of the frozen-out matter in  
central collisions of Au+Au at 4.9 GeV 
energies ($b=$ 2 fm) and Pb+Pb at 17.3 GeV ($b=$ 2.4 fm) 
calculated with the crossover EoS. 
}  
\label{fig-B-Tmu}
\end{figure}

Thus, the freeze-out procedure fixes a single parameter of the matter, 
i.e. the total energy density, that is additionally varied due to 
interference with the fluid dynamics. This results in a whole field of 
temperatures ($T_{\scr{frz}}$) and baryon chemical potentials ($\mu_{\scr{frz}}$) 
of the frozen-out matter in the system. 
To quantify these fields, it is useful to consider distributions of various quantities 
over $T_{\scr{frz}}$ and $\mu_{\scr{frz}}$. In Fig. \ref{fig-B-Tmu} this is done at the 
example of the baryon-charge distribution over the temperature and baryon chemical potential  
of the frozen-out baryon-rich fluids in central collisions at two incident energies, 
$\sqrt{s_{NN}}=$ 4.9 and 17.3 GeV, calculated in the crossover scenario.
As seen, the regions of $T_{\scr{frz}}$ and $\mu_{\scr{frz}}$
are nevertheless well localized rather than extend to the whole available range. 
It should be mentioned that the contribution of rather cold spectator parts of the evolving system 
is excluded in Fig. \ref{fig-B-Tmu}. A weak noise at high $\mu_{\scr{frz}}$ 
illustrates the accuracy of this spectator cutoff.

As has been already mentioned, the model parameters (the friction, the freeze-out energy density and 
the formation time of the fireball fluid) were fitted to reproduce the (net)proton rapidity distributions 
and mid-rapidity pion densities basically at three incident energies $\sqrt{s_{NN}}=$ 4.9,  17.3 and 
62.4\footnote{The results for the  energy of 
62.4 GeV should be taken with care,  
because they are not quite accurate. An accurate computation requires 
unreasonably high memory and CPU time.} 
GeV.  
Though, even with these parameters it was impossible to simultaneously fit all the desired
quantities within the hadronic scenario \cite{Ivanov:2013wha,Ivanov:2013yqa}. 
By means of the above procedure all the model parameters turn out to be determined. 
All other observables, except for those above mentioned, are subjects for predictions of the 
3FD model. It should be mentioned that 
within the deconfinement scenarios the friction in the hadronic phase 
is not a varied quantity but is rather taken from a microscopic estimate of Ref. \cite{Sat90}. 
In fact, there is no need to vary it 
because simulations with the microscopic estimate quite accurately reproduce the data at lower AGS energies. 
In principle, the freeze-out energy density could be fitted separately at each incident energy. 
However, this  gives only a tiny improvement of the data reproduction. 
Therefore, the freeze-out energy density is kept incident-energy independent. 

The phase trajectories presented below were calculated precisely with this parameter set, 
without any additional tuning. The agreement of the 3FD predictions 
with the SM freeze-out points, discussed below, could probably 
be  improved by means of the above-mentioned incident-energy dependent tuning
of the freeze-out energy density. However, this has not been done.

\section{Phase Evolution and Effective Freeze-out}
\label{Phase Evolution}

In the statistical model, mid-rapidity hadron densities are analyzed. 
At high incident energies, longitudinally central and peripheral regions (in space) are 
also well separated in the rapidity space. Therefore, 
only the (spatially) central part the  final freeze-out state predominantly 
contributes to the mid-rapidity density. Thus, it is reasonable to consider evolutions of the matter 
in the central region of the fireball, as it was done in Ref. \cite{Randrup07}. 
Similarly to that it has been done in Ref. \cite{Randrup07}, 
it is useful to study 
trajectories of the matter in the central box placed around the
origin ${\bf r}=(0,0,0)$ in the frame of equal velocities of
colliding nuclei:  $|x|\leq$ 2 fm,  $|y|\leq$ 2 fm and $|z|\leq$
$\gamma_{cm}$ 2 fm, where $\gamma_{cm}$ is Lorentz
factor associated with the initial nuclear motion in the c.m. frame.  
The size of the box was chosen 
to be large enough that the amount of matter in it can be
representative to conclude on properties of the inner part of the system 
and to be small enough to consider the matter in it as a homogeneous
medium.
Contrary to Ref. \cite{Randrup07}, I consider these trajectories 
in terms of temperature ($T$) and baryon chemical potential ($\mu_B$). 
Only expansion stages of the fireball 
evolution are considered because at these stages the system 
is closer to equilibrium than at early stages and hence the above thermodynamic 
quantities are better defined.

Definition of these thermodynamic variables in terms of the 3FD model \cite{3FD}
needs explanations. 
At the expansion stage the baryon-rich fluids in the central region
(i.e. those leading particles which exercised strong stopping) 
are already unified, i.e. mutually stopped and equilibrated,      
while the baryon-free fluid (i.e. the matter produced and predominantly
occupying the central region) is not still equilibrated with the baryon-rich fluids. 
To calculate effective thermodynamic parameters 
of this combined fireball consisting of unified-baryon-rich and baryon-free fluids, 
we have to proceed from its total energy density and baryon density. 
When the two baryon-rich fluids are unified, the calculation of the total baryon density 
is straightforward because the net-baryon charge of the baryon-free fluid is zero. 
The problem occurs with the total energy density. 
In general, the unified baryon-rich fluid and the baryon-free one have different 
local hydro velocities. 
Even if  the total energy density is calculated 
in a local common rest frame of these fluids, a part 
of collective energy associated with the relative hydrodynamic motion of these fluids 
unavoidably gets included in this energy density. This is highly undesirable. 
The only region, where we can safely sum the proper energy densities of 
two discussed fluids, is the central box discussed above. The hydro 
velocities of the two fluids are equal and amount to zero 
(in the c.m. frame of colliding nuclei) for the symmetry reasons.

\begin{figure}[pbht]
\includegraphics[width=8.6cm]{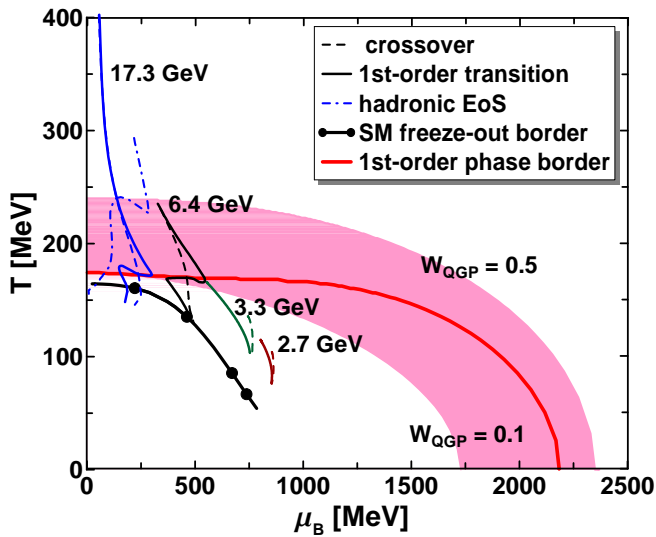}
 \caption{
Phase diagram for the 2-phase EoS (solid thick line) and crossover EoS 
(shaded band) in terms 
of the temperature ($T$) and the baryon chemical potential ($\mu_B$)
and the freeze-out border (solid thick line with dots) deduced from experimental data 
within the statistical model \cite{Andronic:2008gu}.  
For the crossover EoS  the borders of the transition band correspond to
values of the QGP fraction $W_{QGP}=$ 0.1 and $W_{QGP}=$ 0.5.
Dynamical trajectories of the matter in the central box of the
colliding nuclei for three EoS's are also presented.
The trajectories correspond to 
central collisions of Au+Au at energies
$\sqrt{s_{NN}}=$ 2.7 and 3.3 GeV 
 ($b=$ 2 fm) and Pb+Pb at energies 6.4 and 17.3 GeV ($b=$ 2.4 fm). 
Only expansion stages of the
evolution are displayed.  
The freeze-out  points corresponding to displayed collisions 
(the incident energy rises from the right to the left along the freeze-out border)
are taken from 
Ref. \cite{Andronic:2008gu}. 
}  
\label{fig-Tmu}
\end{figure}

Thus, because of the dominant contribution to the mid-rapidity region at high incident energies 
and the possibility of a consistent definition of 
$(T,\mu_B)$ variables of the combined matter, 
the phase-space trajectories of the matter contained in the central box are studied. 
Only central collisions of heavy nuclei are considered: 
Au+Au collisions at impact parameter $b=$ 2 fm
for AGS and RHIC energies,  
and  Pb+Pb collisions at $b=$ 2.4 fm for SPS energies.

In Fig. \ref{fig-Tmu} the phase diagrams for the the 2-phase and crossover EoS's in terms 
of the temperature  and the baryon chemical potential,  
and the freeze-out border deduced from experimental data 
within the statistical model \cite{Andronic:2008gu} are displayed. 
This border and points on it corresponding to specific incident energies 
of central collisions of heavy nuclei are plotted accordingly to 
the  parametrization of the the statistical-model results given in Ref. \cite{Andronic:2008gu}.

In the case of the crossover EoS, only the region of the mixed phase between 
the borders of the QGP fraction of 
$W_{QGP}=$ 0.1 and $W_{QGP}=$ 0.5 is displayed, because in fact the 
hadronic fraction survives up to very high temperatures and chemical potentials.  
In this respect, this version of the crossover EoS certainly contradicts results of the 
lattice QCD calculations, where a fast crossover, at least at zero chemical potential, 
was found \cite{Aoki:2006we}. 
Therefore, a true EoS is somewhere in between  the crossover and 2-phase EoS's 
of Ref. \cite{Toneev06}.

Some examples of trajectories of the matter in the central box are also presented in  
Fig. \ref{fig-Tmu}. 
Only expansion stages of the fireball evolution are displayed. 
The hadronic trajectories are  very close the crossover ones at $\sqrt{s_{NN}}\leq$  6.4 GeV
($E_{lab}\leq$ 20$A$ GeV). Therefore, these are not displayed for the sake of clarity of the figure. 
As seen, only 
comparatively low chemical-potential part of the phase diagram is explored by nuclear collisions. 
At high incident energies, $\sqrt{s_{NN}}\geq$  6.4 GeV ($E_{lab}\geq$ 20$A$ GeV), 
the trajectories quite closely hit 
the corresponding freeze-out points deduced within the statistical model. The exception is the 
hadronic trajectory at $\sqrt{s_{NN}}=$  17.3 GeV ($E_{lab}=$ 158$A$ GeV)  
that ends near the freeze-out border but far from the corresponding 
point. As we will see below, this is a general failure of the hadronic EoS. 

Though the volume of the central box is essentially smaller than that
of the whole system, it is still not always negligible. At $\sqrt{s_{NN}}<$ 9 GeV, 
the freeze-out in the central box occurs practically immediately. With the 
energy rise above 9 GeV, this freeze-out time span becomes nonzero.   
At $\sqrt{s_{NN}}>$ 12 GeV, 
it is of the order of 1 fm/c   
in the c.m. frame of the colliding nuclei.
Therefore, 
the fact that the deconfinement-transition trajectories slightly overshoot the 
SM freeze-out border at  $\sqrt{s_{NN}}=$  17.3 GeV 
is natural because the freeze-out is not immediate.  

The central-box trajectories, 
corresponding to energies 2.7 and 3.3 GeV, 
end sufficiently far from 
the freeze-out border and from the corresponding freeze-out points. 
The reason is that at low energies all spatial parts of the fireball contributes to the 
mid-rapidity region rather than the central part only. Fortunately, at low incident 
energies the baryon-free fluid is underdeveloped. Indeed, this fluid
contributes less than 10\% to the midrapidity value of pions at $\sqrt{s_{NN}}\leq$  3.9 GeV ($E_{lab}\leq$ 6$A$ GeV)
for all considered scenarios, whereas at $\sqrt{s_{NN}}\ge$  6.4 GeV ($E_{lab}\ge$ 20$A$ GeV)
this contribution already amounts to greater than 25\%. Therefore, at $\sqrt{s_{NN}}\lsim$  4 GeV
it is possible to neglect the contribution of the baryon-free fluid that solves the 
problem of the local definition  of the $(T,\mu_B)$ variables in any point of the system 
discussed above.  Thus, it is possible to consider trajectories of the global evolution 
of the system formed from $(T,\mu_B)$ variables averaged over the whole system with the 
weight of the local baryon density of unified baryon fluids. This is done below.

\begin{figure}[pthb]
\includegraphics[width=8.6cm]{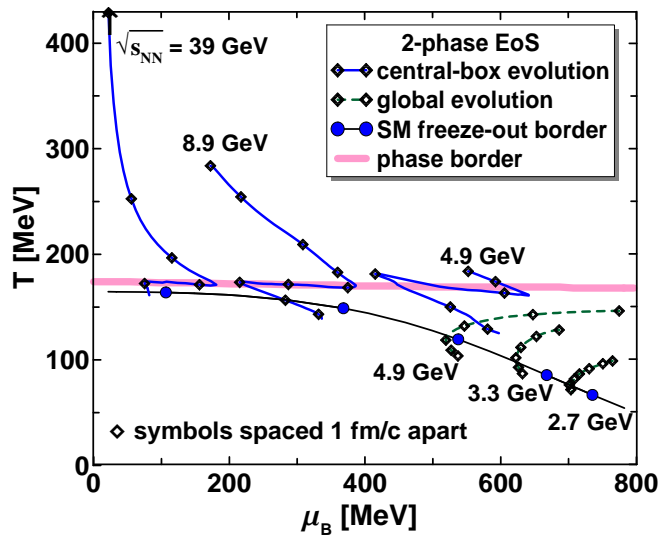}
 \caption{
Dynamical $(T,\mu_B)$-trajectories of the matter in 
central collisions of Au+Au at energies 
$\sqrt{s_{NN}}=$ 2.7, 3.3, 4.9 and  39 GeV 
 ($b=$ 2 fm) and Pb+Pb at  8.9  GeV ($b=$ 2.4 fm) 
calculated with the  2-phase EoS.  
Both the central-box matter (for higher energies) and global (for lower energies) 
evolution trajectories are presented. 
Symbols on the trajectories indicate the time rate of the evolution:
time span between marks is 1 fm/c.  
Only expansion stages of the
evolution are displayed. 
The freeze-out  points correspond to displayed collisions 
(the incident energy rises from the right to the left along the freeze-out  border)
and are taken from the parametrization of the statistical-model results \cite{Andronic:2008gu}. 
} 
\label{fig_Tmu-tph}
\end{figure}
\begin{figure}[pthb]
\includegraphics[width=8.6cm]{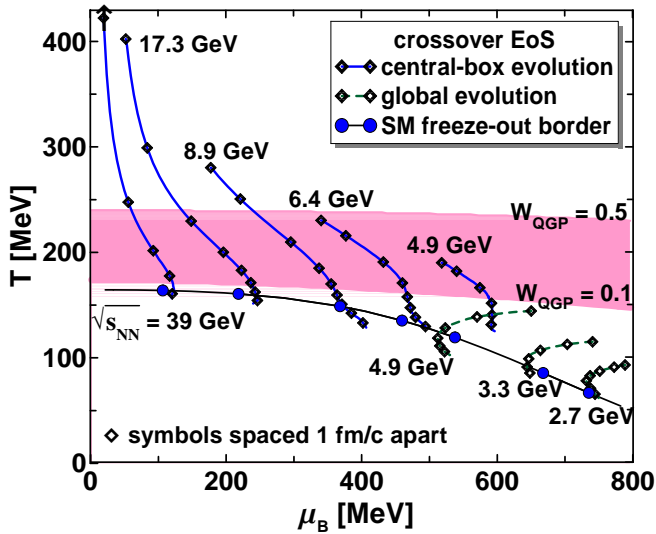}
 \caption{
The same as in Fig. \ref{fig_Tmu-tph} but for the crossover EoS
Trajectories for
central collisions of Au+Au at energies
$\sqrt{s_{NN}}=$ 2.7, 3.3, 4.9 and 39  GeV    
 ($b=$ 2 fm) and Pb+Pb at  6.4, 8.9 and 17.3 GeV ($b=$ 2.4 fm) 
are presented.  
The shadowed ``mixed-phase'' region is located between the borders 
related to
values of the QGP fraction $W_{QGP}=$ 0.1 and $W_{QGP}=$ 0.5.
} 
\label{fig_Tmu-mix}
\end{figure}

In Fig. \ref{fig_Tmu-tph}, it is shown a zoomed part of the 1st-order phase transition, 
which is explored by nuclear collisions. The freeze-out points \cite{Andronic:2008gu} 
correspond to displayed central nuclear collisions. For collisions at high incident energies the 
central-box trajectories are displayed, whereas for lower energies, the trajectories 
of the global evolution. The energy of $\approx$ 5 GeV
is on the border between high and low ones. 
Therefore, both trajectories are presented for this energy. 
These trajectories are very different. The reason for this is the fact that 
the produced excited system is highly nonhomogenous. 
Peripheral regins of the produced fireball are essentially less compressed and 
excited. 
As a result, the form of the trajectory strongly depends on a region over which the averaging runs.
The starting point of the trajectory for $\sqrt{s_{NN}}=$ 39 GeV is beyond the frame 
of Fig. \ref{fig_Tmu-tph},  
it is located at $T\approx 600$ MeV. This fact is 
indicated by the arrow at the top end of this trajectory (it is similarly done in Fig. \ref{fig_Tmu-mix}). 
Symbols mark the time intervals along the trajectory, they are spaced 1 fm/c apart. 
The evolution proceeds from top to bottom of a trajectory.

In Fig. \ref{fig_Tmu-tph} a wiggle characteristic for the 1st-order phase transition
is seen on the trajectories in the region of the transition. 
The length of these wiggles indicate that the central-box matter spends 
a considerable part of the expansion time ($\sim$ 25\%) in the mixed phase. 
The trajectories, the central-box ones at high energies and the global ones   
at low energies, end not far from the corresponding phenomenological 
freeze-out points. The agreement is good while not perfect.

In Fig. \ref{fig_Tmu-mix}, a zoomed part of the crossover transition 
with the matter-evolution trajectories is presented. Here the trajectories
much closer hit the corresponding phenomenological 
freeze-out points than in the case of the first-order-transition scenario. 
Though, this fact does not significantly affect the reproduction of mid-rapidity 
densities of various species \cite{Ivanov:2013yqa}.

\begin{figure}[thb]
\includegraphics[width=8.6cm]{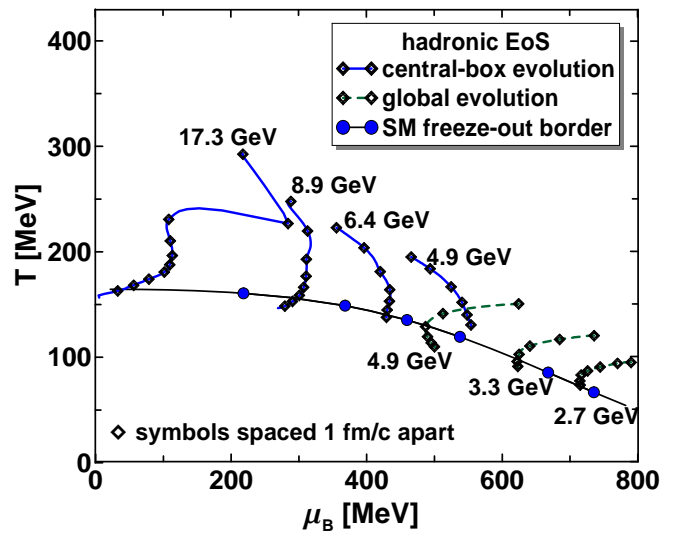}
 \caption{
The same as in Fig. \ref{fig_Tmu-tph} but for the hadronic EoS
Trajectories for
central collisions of Au+Au at 
$\sqrt{s_{NN}}=$ 2.7, 3.3, and 4.9  GeV    
 ($b=$ 2 fm) and Pb+Pb at  6.4, 8.9 and 17.3 GeV ($b=$ 2.4 fm) 
are presented.  
} 
\label{fig_Tmu-gas}
\end{figure}

Fig. \ref{fig_Tmu-gas} presents the phase evolution of the matter within the 
hadronic scenario. 
The wiggle in the hadronic trajectory for $\sqrt{s_{NN}}=$  17.3 GeV
results from the delayed production 
of the baryon-free fluid (i.e. newly produced particles near the mid-rapidity) 
\cite{3FD,Ivanov:2013wha}. This time delay in the hadronic scenario 
amounts to 2 fm/c. Therefore, the baryon-free fluid starts to contribute to the total energy density 
and hence to the effective temperature of the matter only after this time span. 
Naturally it raises the temperature. When the baryon-free fluid got completely
formed, the trajectory returns to it natural behavior.  Such a wiggle is absent on the 
trajectories related to 2-phase and crossover EoS's because in those cases the delay time 
amounts to 0.17 fm/c and hence the formation of the baryon-free fluid gets completed already 
at the compression stage of the collision. 
The delay time for each scenario was chosen proceeding from the best 
reproduction of available experimental 
data. Notice that at high incident energies, 
when the baryon-free fluid is already well developed, 
 the delay time essentially affects the baryon stopping, at $\sqrt{s_{NN}}=$ 39 GeV 
 it even becomes decisive. The earlier the baryon-free fluid 
 is produced, the earlier it starts to interact with baryonic fluids and hence the 
 stronger baryon stopping provides.

In the case of the hadronic scenario the agreement with the corresponding phenomenological 
freeze-out points is the worst among the considered scenarios even at low incident energies, 
where a pure hadronic dynamics takes place.  
Probably the latter is a byproduct of enhancement 
the inter-fluid friction in the hadronic phase \cite{Ivanov:2013wha,3FD} 
as compared with its microscopic estimate 
of Ref. \cite{Sat90}. This enhancement has been applied in order to reproduce a
major part (however not all \cite{Ivanov:2013yqa}) of observables up to the energy of 17.3 GeV
This modification of the friction spoils 
the agreement in the purely hadronic domain. 
The advantage of  deconfinement-transition scenarios is that they do not require any modification 
of the microscopic friction in the hadronic phase.

\section{Summary}

Evolution of the matter in relativistic collisions of heavy nuclei
and the resulting  freeze-out parameters  
were analyzed   
 in the incident energy range of 2.7 GeV $\le\sqrt{s_{NN}}\le$ 39 GeV.  
These simulations were performed within the 3FD model  
\cite{3FD} employing three different equations of state: a purely hadronic EoS   
\cite{gasEOS}, and two versions of EoS involving the deconfinement 
 transition \cite{Toneev06}, i.e. an EoS with the first-order phase transition
and that with a smooth crossover transition. 
Details of these calculations are described in Ref. \cite{Ivanov:2013wha}.

It is found that the freeze-out parameters deduced from experimental data 
within the statistical model \cite{Andronic:2008gu} are well reproduced within the crossover scenario. 
The 1st-order-transition scenario turns out to be slightly less successful. 
In the case of the hadronic scenario the agreement with the corresponding phenomenological 
freeze-out points is the worst among the considered scenarios even at low incident energies, 
where a pure hadronic dynamics takes place.  
Probably the latter is a byproduct of noticeable enhancement 
the inter-fluid friction in the hadronic phase \cite{Ivanov:2013wha,3FD} 
as compared with its microscopic estimate 
of Ref. \cite{Sat90}, that was introduced in order to reproduce a 
major part (however not all \cite{Ivanov:2013yqa}) of observables up to the energy of 17.3 GeV

In particular, these results explain why 
the hadronic scenario fails to reproduce 
experimental yields of antibaryons (strange and nonstrange), 
starting already from lower SPS energies, i.e. $\sqrt{s_{NN}}\le$ 6.4 GeV, 
and  yields of all other species
at energies above the top SPS one, i.e. $\sqrt{s_{NN}}>$ 17.3 GeV, while 
the  deconfinement-transition scenarios reasonably agree (to a various extent) 
with all the data \cite{Ivanov:2013yqa}.

The present analysis, as well as results of Ref. \cite{Ivanov:2013yqa} indicates 
a certain preference of the deconfinement-transition  EoS 
which predict onset of the deconfinement in central collisions of heavy nuclei 
at top AGS energies, i.e. $\sqrt{s_{NN}}\gsim$ 5 GeV. 
However, it should be mentioned that the crossover transition constructed in Ref. \cite{Toneev06}
is very smooth  \cite{Ivanov:2012bh,Ivanov:2013wha}.  
In this respect, this version of the crossover EoS certainly contradicts results of the 
lattice QCD calculations, where a fast crossover, at least at zero chemical potential, 
was found \cite{Aoki:2006we}. 
Therefore, for better reproduction of experimental data and phenomenological freeze-out 
parameters a more realistic EoS is required. 
\\

I am grateful to A.S. Khvorostukhin, V.V. Skokov,  and V.D. Toneev for providing 
me with the tabulated 2-phase and crossover EoS's. 
The calculations were performed at the computer cluster of GSI (Darmstadt). 
This work was supported by The Foundation for Internet Development (Moscow)
and also partially supported  by  
the grant NS-215.2012.2.

\end{document}